\documentclass[twoside,twocolumn,9pt]{article}
\usepackage{extsizes}

\usepackage[super,sort&compress,comma]{natbib}
\usepackage[version=4]{mhchem}
\usepackage[left=1.5cm, right=1.5cm, top=1.785cm, bottom=2.0cm]{geometry}
\usepackage{balance}
\usepackage{mathptmx}
\usepackage{sectsty}
\usepackage{graphicx} 
\usepackage{lastpage}
\usepackage[format=plain,justification=justified,singlelinecheck=false,font={stretch=1.125,small,sf},labelfont=bf,labelsep=space]{caption}
\usepackage{float}
\usepackage{fancyhdr}
\usepackage{fnpos}
\usepackage[english]{babel}
\addto{\captionsenglish}{%
  
}
\usepackage{array}
\usepackage{droidsans}
\usepackage{charter}
\usepackage[T1]{fontenc}
\usepackage[usenames,dvipsnames]{xcolor}
\usepackage{setspace}
\usepackage[compact]{titlesec}
\usepackage{hyperref}
\usepackage{pdfpages}
\usepackage{multirow}

\usepackage{epstopdf}

\definecolor{cream}{RGB}{222,217,201}

\usepackage{booktabs}       
\usepackage{subcaption}
\usepackage[list-units=single,detect-all]{siunitx}
\DeclareSIUnit{\calorie}{cal}
\DeclareSIUnit{\kcal}{\kilo\calorie\per\mol}
\DeclareSIUnit{\angstrom}{\text {Å}}
\DeclareSIUnit{\kjoule}{\kJ\per\mol}

\usepackage{xr}
\begin{document}

\makeFNbottom
\makeatletter
\renewcommand\LARGE{\@setfontsize\LARGE{15pt}{17}}
\renewcommand\Large{\@setfontsize\Large{12pt}{14}}
\renewcommand\large{\@setfontsize\large{10pt}{12}}
\renewcommand\footnotesize{\@setfontsize\footnotesize{7pt}{10}}
\makeatother

\renewcommand{\thefootnote}{\fnsymbol{footnote}}
\renewcommand\footnoterule{\vspace*{1pt}%
\color{cream}\hrule width 3.5in height 0.4pt \color{black}\vspace*{5pt}} 
\setcounter{secnumdepth}{5}

\makeatletter 
\renewcommand\@biblabel[1]{#1}            
\renewcommand\@makefntext[1]%
{\noindent\makebox[0pt][r]{\@thefnmark\,}#1}
\makeatother 
\renewcommand{\figurename}{\small{Fig.}~}
\sectionfont{\sffamily\Large}
\subsectionfont{\normalsize}
\subsubsectionfont{\bf}
\setstretch{1.125} 
\setlength{\skip\footins}{0.8cm}
\setlength{\footnotesep}{0.25cm}
\setlength{\jot}{10pt}
\titlespacing*{\section}{0pt}{4pt}{4pt}
\titlespacing*{\subsection}{0pt}{15pt}{1pt}

\newcommand{\etal}{\textit{et al.}}

\twocolumn[
  \begin{@twocolumnfalse}

\begin{center}
\LARGE{\textbf{Is the protactinium(V) mono-oxo bond weaker than what we thought?$^\dag$}}
\end{center}

\vspace{0.3cm}

\begin{center}
\large{Tamara Shaaban,\textit{$^{a}$} Hanna Oher,\textit{$^{b}$} Jean Aupiais,\textit{$^{c}$} Julie Champion,\textit{$^{d}$} André Severo Pereira Gomes,\textit{$^{a}$} Claire Le Naour,\textit{$^{b}$} Melody Maloubier,\textit{$^{b}$} Florent Réal,\textit{$^{a}$} Eric Renault,\textit{$^{e}$}  Xavier Rocquefelte,\textit{$^{f}$}  Bruno Siberchicot,\textit{$^{c}$} Valérie Vallet,\textit{$^{a}$} and Rémi Maurice$^{\ast}$\textit{$^{f}$}}
\end{center}

\vspace{0.3cm}

\noindent\normalsize{The bond distance is the simplest and most obvious indicator of the nature of a given chemical bond. However, for rare chemistry, it may happen that it is not yet firmly established. In this communication, we will show that the formally-triple protactinium(V) mono-oxo bond is predicted longer than what was previously reported in the solid state and in solution, based on robust quantum mechanical calculations, supported by an extensive methodological study. Furthermore, additional calculations are used to demonstrate that the \ce{Pa-O_{oxo}} bond of interest is more sensitive to complexation than the supposedly analogous \ce{U-O_{yl}} ones, not only in terms of bond distance but also of finer bond descriptors associated with the effective bond multiplicity.} 

 \end{@twocolumnfalse} \vspace{0.3cm}

\vspace{0.6cm}

]

\footnotetext{\textit{$^{a}$~Univ. Lille, CNRS, UMR 8523 - PhLAM - Physique des Lasers Atomes et Molécules, F-59000 Lille, France}}
\footnotetext{\textit{$^{b}$~Université Paris-Saclay, CNRS/IN2P3, IJCLab, 91405 Orsay, France. }}
\footnotetext{\textit{$^{c}$~CEA, Laboratoire Matière en Conditions Extrêmes, Université Paris-Saclay,
F-91680, Bruyères-le-Châtel, France; CEA, DAM, DIF, 91297 Arpajon, France}}
\footnotetext{\textit{$^{d}$~IMT Atlantique, Nantes Université, CNRS/IN2P3, SUBATECH, F-44000 Nantes, France}}
\footnotetext{\textit{$^{e}$~Nantes Université, CNRS, CEISAM UMR 6230, F-44000 Nantes, France}}
\footnotetext{\textit{$^{f}$~Univ Rennes, CNRS, ISCR (Institut des Sciences Chimiques de Rennes) -- UMR 6226, F-35000 Rennes, France. E-mail: remi.maurice@univ-rennes.fr}}
\footnotetext{\dag~Electronic Supplementary Information (ESI) available: Additional calculations essentially of interest for methodological purposes.}

Protactinium (Pa, $Z$ = 91) is an actinide with enigmatic physico-chemical properties \cite{Pa, WilsonNC} that would differentiate the ``transition-metal'' like actinides (up to thorium, Th) from the heavier ones (starting with uranium, U) \cite{ValletPa}. One of the key features of protactinum is its propensity to form a single oxo bond in solution in its pentavalent oxidation state\cite{LeNaour:2022}, while it is not observed for the ``transition-metal'' like actinides and while dioxo bonds are observed for the heavier elements such as uranium, neptunium (Np), plutonium (Pu) and americium (Am). 

This single oxo bond has been characterized by X-ray absorption spectroscopy in a couple of systems\cite{Mendes:2010, LeNaour:2005}, in particular \ce{[PaO(C2O4)3]^{3-}} and \ce{[PaO(SO4)3]^{3-}}, with a fitted \ce{Pa-O_{oxo}} bond distance of $\sim$\qtyrange[range-phrase=--]{1.72}{1.75}{\angstrom}, depending on the medium and conditions. The objective of the present article is to re-evaluate this bond distance from good theoretical grounds, by going well beyond our previous articles on molecular complexes that already pointed out a potential discrepancy between theory and experiment \cite{PaO-ox3, Chem-2024}. To be free of any ambiguity related to speciation in solution, we include in our study the sole solid-state compound ever reported with an occurring \ce{Pa-O_{oxo}} bond\cite{CrystalS}, namely \ce{[C8H_{20}N]2[PaOCl5]}. However, we must stress that post-Hartree-Fock methods are not generally available in periodic codes, which is why we have also performed a detailed methodological study on both model and actual molecular systems (\textit{vide infra}). 

Experimental lattice parameters and bond distances are reported in \autoref{tab:1}. In this structure, the \ce{[PaOCl5]^{2-}} molecular unit emerges, with one shorter axial \ce{{Pa--Cl}_{ax}} bond distance and four longer equatorial \ce{{Pa--Cl}_{eq}} bond distances (non equivalent in the reported $Cc$ space group), together with a \ce{Pa-O_{oxo}} bond distance of interest of \qty{1.74(9)}{\angstrom}. Periodic density functional theory (DFT) calculations were performed with the VASP code\cite{VASP, PAW}. Both lattice parameters and atomic positions were optimized with various setups (see~\autoref{tab:1}). For all reported calculations, the same plane-wave energy cutoff of \qty{550}{\electronvolt} was applied. From the generic PBE\cite{PBE} calculation, the f levels can be effectively pushed up by an \textit{ad hoc} Hubbard U term\cite{U} or finer described by using a hybrid exchange-correlation functional such as PBE0\cite{PBE0}, and on top of that a (post-SCF) dispersion correction may be applied\cite{D3BJ}. 

It is clear that for obtaining a fair agreement between the computed and experimental lattice parameters the introduction of a dispersion correction, namely D3(BJ)\cite{D3BJ}, is required (see \autoref{tab:1}). Once it is applied, both the PBE and PBE0 functionals yield similar lattice parameters. While the \ce{Pa-O_{oxo}} bond distance is typically increased by the +U correction, the introduction of Hartree-Fock exchange reduces it. Note that the dispersion correction keeps this bond distance untouched, with both the PBE and PBE0 functionals. With the more accurate PBE0 functional, we obtain a \ce{Pa-O_{oxo}} bond distance of \qty{1.83}{\angstrom}, which lies within the confidence interval of the experimental distance, 1.74$\pm$0.09 {\AA}. 

\begin{table*}
\centering
\small
\caption{Optimized lattice parameters and relevant bond distances obtained for \ce{[C8H_{20}N]2[PaOCl5]} by periodic density functional theory calculations. Results are compared with experiment ($Cc$ space group).}   
\label{tab:1}
\setlength{\tabcolsep}{3pt}
\begin{tabular*}{0.96\textwidth}{@{\extracolsep{\fill}}l*{7}l}
\hline
\ce{[C8H_{20}N]2[PaOCl5]} & {$a$ (\AA)} & {$b$ (\AA)} & {$c$ (\AA)} & {$\beta$ ($^\circ$)} & {$d[\text{Pa--O}]$ (\AA)} & {$d[\ce{{Pa--Cl}_{ax}}]$ (\AA)} & {$\overline{d}[\ce{{Pa--Cl}_{eq}}]$ (\AA)$^a$} \\
\hline
PBE\cite{PBE} & 14.281 & 14.951 & 13.596 & 91.34 & 1.859 & 2.648 & 2.689(6) \\
PBE+D3(BJ)\cite{PBE, D3BJ} & 13.301 & 14.840 & 12.745 & 91.65 & 1.859 & 2.624 & 2.673(17) \\
PBE+U+D3(BJ)\cite{PBE, D3BJ, U} (U$_\text{eff}$ = 4 eV) & 13.295 & 14.797 & 12.825 & 91.36 & 1.890 & 2.631 & 2.690(14) \\
\hline
PBE0\cite{PBE0} & 13.996 & 14.799 & 13.350 & 91.29 & 1.832 & 2.624 & 2.674(7) \\
PBE0+D3(BJ)\cite{PBE0, D3BJ} & 13.153 & 14.665 & 12.518 & 91.59 & 1.832 & 2.608 & 2.659(19) \\
\hline
Expt.\cite{CrystalS} & 14.131(8) & 14.218(8) & 13.235(9) & 91.04(3) & 1.74(9) & 2.42(3) & 2.64(6) \\
\hline
\end{tabular*}
\flushleft $^a$The four distances are crystallographically independent.
\end{table*} 

Given the computational cost and for comparison purposes, we retained the PBE+D3(BJ) method\cite{PBE,D3BJ} for probing two well-resolved, recent, uranium structures\cite{AngewS, ICS}, namely \ce{(C4H_{12}N2)[UO2Cl4]} and \ce{(C4H_{12}N2)2[UO2Cl4(H2O)]Cl2}. We note that a very good agreement with experiment is observed for the lattice parameters and for the U--Cl bond distances (see~\autoref{tab:2}), which validates the chosen level of theory. Concerning the \ce{U-O_{yl}} dioxo bond distances, we observe a standard overestimation of $\sim$\qtyrange{0.04}{0.05}{\angstrom} at this level of theory. If we correct the previous PBE+D3(BJ) \ce{Pa-O_{oxo}} bond distance by the opposite amount, we obtain $\sim$\qtyrange{1.81}{1.82}{\angstrom} , which again lies within the experimental confidence interval. We thus conclude at this stage that the \ce{Pa-O_{oxo}} bond distance is predicted to be slightly longer than \qty{1.8}{\angstrom} and thus also longer than the typical \ce{U-O_{yl}} ones.\cite{Chem-2024}

\begin{figure}[htbp]
\includegraphics[scale=0.45]{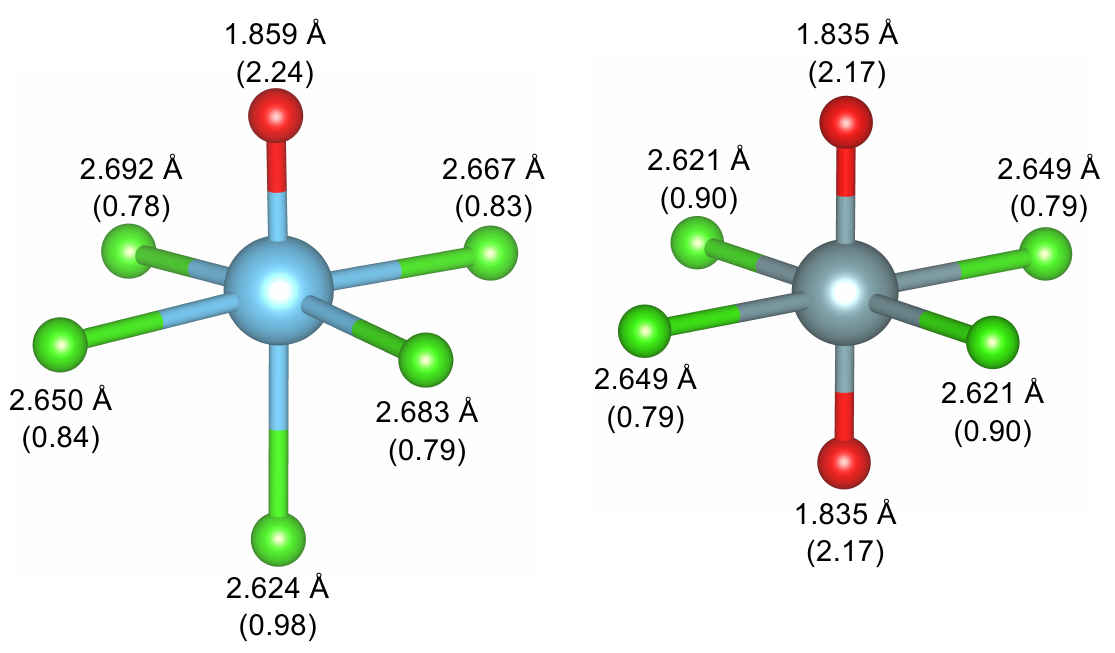}
\caption{\label{structure} Schematic representation of the PBE+D3(BJ) atomic arrangements of the \ce{[PaOCl5]^{2-}} (left) and \ce{[UO2Cl4]^{2-}} (right) units (visualization performed with VESTA \cite{Vesta}). The ICOBIs are given in brackets.}
\end{figure}

Without entering yet in a detailed discussion concerning the \ce{Pa-O_{oxo}} bond nature, we have performed additional periodic calculations with the PBE functional for the bare \ce{[PaO]^{3+}} and \ce{[UO2]^{2+}} systems as well as determined integrated crystal orbital bond indices (ICOBIs)\cite{ICOBIs} at the PBE+D3(BJ) level on the previous solid-state systems (see text below \autoref{tab:S1}). Note that the reference bonds in \ce{[PaO]^{3+}} and \ce{[UO2]^{2+}} are both formally triple\cite{Chem-2024}, with a subtle distinction: in \ce{[PaO]^{3+}} bonding orbitals may display mixed 6d/5f Pa characters, whereas in \ce{[UO2]^{2+}}, the 6d and 5f U characters are symmetry separated. In the bare systems, the \ce{Pa-O_{oxo}} bond is slightly longer and displays a higher ICOBI. This is also true for the  \ce{[C8H_{20}N]2[PaOCl5]} and \ce{(C4H_{12}N2)[UO2Cl4]} solid-state systems (see ~\autoref{structure}). Naturally, the ICOBIs are there smaller than in the bare systems (longer distances). We note that the full crystal environment induces a lengthening of the \ce{Pa-O_{oxo}} bond by \qty{0.037}{\angstrom}, larger than that found in uranyl, \qty{0.024}{\angstrom}. This, plus the shorter \ce{{Pa--Cl}_{ax}} bond distance indicates that the \ce{Pa-O_{oxo}} bond is more sensitive to complexation since the \textit{trans} position is not protected, unlike in \ce{[UO2]^{2+}} and related compounds.

\begin{table*}
\centering
\small
\caption{Optimized lattice parameters and relevant bond distances obtained for \ce{(C4H_{12}N2)[UO2Cl4]} and \ce{(C4H_{12}N2)2[UO2Cl4(H2O)]Cl2}, two representative uranyl compounds, by periodic density functional theory calculations. Results are compared with experiment ($P\overline{1}$ and $Pnma$ space groups).}   
\label{tab:2}
\setlength{\tabcolsep}{3pt}
\begin{tabular*}{0.96\textwidth}{@{\extracolsep{\fill}}l*{8}c}
\hline
\ce{(C4H_{12}N2)[UO2Cl4]} & {$a$ (\AA)} & {$b$ (\AA)} & {$c$ (\AA)} & {$\alpha$ ($^\circ$)} & {$\beta$ ($^\circ$)} & {$\gamma$ ($^\circ$)} & {$d[\text{U--O}]$ (\AA)} & {$\overline{d}[\text{U--Cl}]$ (\AA)$^a$} \\
\hline
PBE+D3(BJ)\cite{PBE, D3BJ} & 6.668  & 6.790 & 7.300 & 81.80 & 85.80 & 63.36 & 1.835 & 2.635(14) \\
(dev.) & (+0.75\%) & (+1.94\%) & ($-$2.12\%) & ($-$1.80\%) & ($-$2.17\%) & (+0.14\%) \\
Expt.\cite{AngewS} & 6.6183(6) & 6.6609(6) & 7.4578(6) & 83.300(3) & 87.703(3) & 63.273(2) & 1.786(3) & 2.646(9) \\
\hline
\ce{(C4H_{12}N2)2[UO2Cl4(H2O)]Cl2} & {$a$ (\AA)} & {$b$ (\AA)} & {$c$ (\AA)} & & & {$d[\text{U--O}_\text{w}]$ (\AA)} & {$\overline{d}[\text{U--O}_\text{yl}]$ (\AA)$^b$}& {$\overline{d}[\text{U--Cl}]$ (\AA)$^a$} \\
\hline
PBE+D3(BJ)\cite{PBE, D3BJ} & 12.172 & 12.418 & 13.409 & & & 2.513 & 1.8035(5) & 2.768(3) \\
(dev.) & ($-$0.34\%) & ($-$0.22\%) & ($-$0.36\%) \\
Expt.\cite{ICS} & 12.2130(7) & 12.4456(7) & 13.4579(8) & & & 2.499(3) &  1.761(1) & 2.765(3)\\
\hline
\end{tabular*}
\flushleft $^a$The four distances are only crystallography equivalent two by two. \hfill $^b$The two distances are crystallography independent.
\end{table*}

We now move on with complexes in solution. Static calculations, based on electronic structure theory and the application of a continuum dielectric model for solvation (implicit solvation model), lead to longer \ce{Pa-O_{oxo}} bond distances\cite{Mendes:2010, PaO-ox3} than what was reported by EXAFS in solution, by $\sim$0.10-0.15 {\AA}, \textit{i.e.} by more than the expected sum of the experimental and computational errors. Since this discrepancy can originate from the relative simplicity of the model used to mimic the complex in solution, we have chosen to initiate our analysis by comparing results from static quantum chemical (QC) calculations (with isolated systems) and QC molecular dynamics (MD) simulations (with systems embedded in large simulation boxes containing around 100 water molecules). Calculations and simulations were performed with the Abinit code\cite{Abinit}. For this part, we selected two prototypes: \ce{PaOF_3(H2O)3}, previously included in an \textit{ab initio} molecular dynamics (AIMD) study\cite{Fluoro}, and the \ce{[PaO(C2O4)3]^{3-}} complex, previously studied by static approaches\cite{Mendes:2010, PaO-ox3}. In both cases, the protactinium(V) ion displays a coordination number of 7, which practically seems to be the largest possible one in solution for this ion\cite{Fluoro, PaO-ox3}, except in the peculiar \ce{[PaF8]^{3-}} case\cite{PaF8}.

First, we tested ten \textit{pure} (\textit{i.e.} nonhybrid) exchange-correlation functionals (see~\autoref{tab:S2}). For the bare \ce{[PaO]^{3+}} system as well as for the isolated \ce{PaOF_3(H2O)3} and \ce{[PaO(C2O4)3]^{3-}} complexes, we found very homogeneous results for the \ce{Pa-O_{oxo}} bond distance of interest. Therefore, any of those ten functionals could have been used for the AIMD study. We selected the PBE functional\cite{PBE} for the sake of homogeneity across our reported results. As can be seen in~\autoref{tab:3}, the combined influence of the water bath and molecular dynamics has minimal impact on the \ce{Pa-O_{oxo}} bond distance. More importantly, these effects together slightly increase the bond distance (by \qtyrange{0.003}{0.004}{\angstrom}). Therefore, the previously mentioned discrepancy between theory and EXAFS on \ce{[PaO(C2O4)3]^{3-}} cannot be attributed to the static approach. Consequently, to facilitate a more in-depth methodological study, we will now exclusively use the static QC approach.

\begin{table}
\centering
\small
\caption{Joint effect of explicit solvation and molecular dynamics on the Pa--O bond distance (\AA) obtained with the Abinit code\cite{Abinit} and the PBE exchange-correlation functional\cite{PBE}.}   
\label{tab:3}
\setlength{\tabcolsep}{3pt}
\begin{tabular}{@{\extracolsep{\fill}}l*{2}{S[table-format=1.3]}}
\hline
Computational setup & {\ce{PaOF3(H2O)3}} & {\ce{[PaO(C2O4)3]^{3-}}} \\
\hline
Static, isolated molecule & 1.833 & 1.869 \\
Dynamic (298 K), water bath, 1.5 ps & 1.836 & 1.873 \\
\hline
\end{tabular}
\end{table}

We have selected two protactinium(V) systems for this static study, namely \ce{[PaO]^{3+}}, for obvious reasons, and \ce{[PaO(C2O4)3]^{3-}} as our primary case in solution. The sulfuric case\cite{LeNaour:2005} is intentionally excluded due to the unclear coordination modes of the sulfate groups (monodentate \textit{vs.} bidendate fashions), though a similar issue arises for the  \ce{Pa-O_{oxo}} bond distance \cite{PaO-ox3}. Calculations were performed with three computational chemistry codes, namely Gaussian\cite{Gaussian}, Molpro\cite{Molpro} and ADF \cite{ADF}. 

Since detailed results are reported in the Supporting Information (see in particular \autoref{tab:S3} and \autoref{tab:s4}), we only highlight our main conclusions here. First, both employed implicit solvation models lead to a slight \ce{Pa-O_{oxo}} bond lengthening of less than \qty{0.01}{\angstrom} in the \ce{[PaO(C2O4)3]^{3-}} complex, which has a saturated protactinium(V) coordination sphere. Second, the spin-orbit coupling leads to an even more negligible \ce{Pa-O_{oxo}} bond lengthening. Therefore, we report either scalar or spin-orbit relativistic values for comparison purposes, but to keep things simple, we have retained scalar relativistic values in~\autoref{tab:4}. Third, we confirm that by going from PBE\cite{PBE} to the hybrid PBE0\cite{PBE0} functional, the \ce{Pa-O_{oxo}} bond distance is reduced by $\sim$\qty{0.03}{\angstrom}, in accord with the already discussed solid-state results (\textit{vide supra}). Fourth, hybrid functionals such as B3LYP\cite{B3LYP} and PBE0\cite{PBE0} show good agreement with the higher level CCSD(T) method\cite{CCSD(T)}, as does MP2\cite{MP2}. Last, the application of a dispersion correction\cite{D3BJ} is bland on the \ce{Pa-O_{oxo}} bond distance. All-in-all, we have reported the COSMO\cite{COSMO-ADF} scalar relativistic PBE0\cite{PBE0} results in~\autoref{tab:4}.

\begin{table}
\centering
\small
\caption{Scalar-relativistic An--O (An = Pa, U) bond distances obtained with the ADF code\cite{ADF}, the COSMO implicit solvation\cite{COSMO-ADF} and the PBE0 exchange-correlation functional\cite{PBE0}, together with the corresponding delocalisation indices ($\delta$'s)\cite{DI} arising from the QTAIM\cite{QTAIM} analysis.}   
\label{tab:4}
\setlength{\tabcolsep}{3pt}
\begin{tabular}{@{\extracolsep{\fill}}lS[table-format=1.5(1)]S[table-format=1.2]}
\hline
Molecular system & {$d$ (\AA)} & {$\delta$} \\
\hline
\ce{[PaO]^{3+}} & 1.710 & 2.33 \\
\ce{[PaO(C2O4)3]^{3-}} & 1.858 & 1.69 \\
\hline
\ce{[UO2]^{2+}} & 1.702 & 2.15 \\
\ce{[UO2Cl4]^{2-}} & 1.766 & 1.89 \\
\ce{[UO2(C2O4)3]^{4-}} & 1.7815(15) & 1.84\\
\hline
\end{tabular}
\end{table}

Before commenting on the obtained results with protactinium(V), we first assess the quality of the results obtained at the chosen level of theory based on uranyl systems, namely \ce{[UO2]^{2+}}, the well-known \ce{[UO2Cl4]^{2-}} molecular unit (already present in the \ce{(C4H_{12}N2)[UO2Cl4]} solid-state system), and the \ce{[UO2(C2O4)3]^{4-}} complex. The latter has been studied both experimentally and computationally\cite{UO2-ox3} and notably exhibits a uranium(VI) coordination number of 7, making it analogous to the \ce{[PaO(C2O4)3]^{3-}} case. As shown in~\autoref{tab:s5}, the influence of the spin-orbit coupling on the \ce{U-O_{yl}} bond distance(s) is minimal, justifying our selection of a scalar relativistic level in~\autoref{tab:4}. Note that the PBE0 gas-phase result for \ce{[UO2]^{2+}} (\qty{1.68}{\angstrom}) is in fair agreement with the reference \qty{1.71}{\angstrom} value of \citet{deJong:1999}. For \ce{[UO2Cl4]^{2-}} and \ce{[UO2(C2O4)3]^{4-}}, our \ce{U-O_{yl}} bond distance values in solution of \qtylist{1.77; 1.78}{\angstrom}, respectively, are furthermore of quite exceptional match with the \qtylist{1.77(1); 1.79(4)}{\angstrom} experimental ones\cite{UO2Cl4, UO2-ox3}. Thus, we conclude that the values reported in~\autoref{tab:4} are of sufficient quality to allow for reliable comparisons between the protactinium(V) and uranium(VI) cases.

From~\autoref{tab:4}, it is evident that while the \ce{An-O} bond distance in the bare \ce{[PaO]^{3+}} system is slightly longer than that in \ce{[UO2]^{2+}}, it is much longer in \ce{[PaO(C2O4)3]^{3-}} than in \ce{[UO2Cl4]^{2-}} and \ce{[UO2(C2O4)3]^{4-}}. In other words, even if \ce{[PaO]^{3+}} may be seen as a ``half'' \ce{[UO2]^{2+}} unit, the coordination-induced bond lengthening is much more pronounced in the case of protactinium(V) (\textit{ca.}~\qty{+0.15}{\angstrom}) than in the uranium(VI) one (\textit{ca.}~\qty{+0.07}{\angstrom}), being in practice doubled. Thus, we definitely conclude that \ce{[PaO]^{3+}} exhibits a much stronger sensitivity to extra coordination compared to \ce{[UO2]^{2+}}.

To further check if this perspective on bond distance corroborates with the bonding perspective, we have also determined delocalisation indices ($\delta$'s)\cite{DI} within the QTAIM framework\cite{QTAIM} (other indicators are given in Table \ref{tab:s6}). We recall that $\delta$ values, by definition, correspond to the number of electron pairs shared between two atoms, reflecting the effective bond multiplicity. In the bare \ce{[UO2]^{2+}} system, six bonding orbitals of $\sigma_{u,g}$ and $\pi_{u,g}$ are doubly occupied in the ground state\cite{Denning}, the \textit{ungerade} ones involving 5f orbitals of U and the \textit{gerade} ones involving 6d orbitals of U. Thus, the \ce{U-O_{yl}} bonds are formally triple bonds. As shown in~\autoref{tab:4}, $\delta$ in there much smaller than 3, which is due to the strong asymmetry of those bonds (``polar'' character\cite{PolarBond}). Upon complexation, $\delta$ is reduced by $\sim$0.3 in both \ce{[UO2Cl4]^{2-}} and \ce{[UO2(C2O4)3]^{4-}}. Note that our 1.89 value for \ce{[UO2Cl4]^{2-}} is consistent with the 1.87 one of \citet{PolarBond}.

In the bare \ce{[PaO]^{3+}} system, three bonding orbitals are populated in the ground state. Note that due to the absence of a symmetry center, the 5f and 6d orbitals of Pa are not symmetry separated anymore. The obtained $\delta$ value, 2.33, is slightly higher than in \ce{[UO2]^{2+}}, which is 2.15. This means that the slightly longer \ce{Pa-O_{oxo}} bond distance goes together with a slightly more covalent \ce{Pa-O_{oxo}} bond in \ce{[PaO]^{3+}} compared to the \ce{U-O_{yl}} ones in \ce{[UO2]^{2+}}. This is attributed to the smaller formal charge of protactinium(V) \textit{versus} uranium(VI), hence a larger ionic radius. After complexation, $\delta$ is reduced by $\sim$0.6 in \ce{[PaO(C2O4)3]^{3-}}. It is interesting to note that the observed factor of 2 in the complexation-induced bond lengthening perfectly correlates with the factor of 2 reduction in $\delta$. Thus, these two bond descriptors point toward the exact same conclusion. 

In this article, we have shown that the \ce{Pa-O_{oxo}} bond is more sensitive to complexation than the \ce{U-O_{yl}} ones. We attribute this to the fact that the \textit{trans} position is not protected in the \ce{[PaO]^{3+}} chemical moiety. This leaves room for coordination with specific bonding at this position in real chemical systems. For instance, it can occur in the solid-state \ce{[PaOCl5]^{2-}} molecular unit (see Tables \ref{tab:s7}, \ref{tab:s8} and \ref{tab:s9} and the associated discussion on bonding) or in the \ce{[PaO(C2O4)3]^{3-}} complex. Specifically, the distance to the \textit{trans} ligands is expected to be shorter than to the \textit{cis} ones due to the inverse \textit{trans} effect already observed and analyzed in actinide complexes~\cite{actinide-Pershina-RA1999-84-79--84,actinide-Pershina-RA1998-80-65--73,actinide-OGrady-JCDT2002--1233--1239,Fryer-Kanssen-Chem.Commun.-2018-54-9761-9764,actinide-Motta-NC2023-14-4307}. Since the \ce{Pa-O_{oxo}} bond distance is predicted to be longer than the \ce{U-O_{yl}} one, the previous experimentally reported \ce{Pa-O_{oxo}} bond distances must be revised. According to all of our calculations, this bond distance is in any case predicted to be slightly longer than \qty{1.8}{\angstrom}, indicating that it is weaker than previously thought. An obvious perspective of this work would consist in performing new experiments to confirm these findings. Furthermore, this work is already a key step in the exploration of the intriguing physico-chemical properties of the actinides since it confirms the peculiarity of protactinium, which definitely qualifies it as as a trend separator within the actinide series.

\section*{Author Contributions}
This article is the result of a collective effort supported by the ANR CHESS project (contract No. ANR-21-CE29-0027). T.S. is the primary contributor for the performed calculations, completed by additional work by H.O., X.R. and B.S. The first draft was written by R.M. and each version was approved by all the co-authors.

\section*{Data availability}
The data supporting this article have been included as part of the Supplementary Information$^\dag$ and the optimized structures can be accessed at the Zenodo repository via the following DOI: \href{https://doi.org/10.5281/zenodo.13494221}{10.5281/zenodo.13494221}.

\section*{Conflicts of interest}
There are no conflicts to declare.





\scriptsize{
\bibliographystyle{rsc} 
\bibliography{pao-bond}

\providecommand*{\mcitethebibliography}{\thebibliography}
\csname @ifundefined\endcsname{endmcitethebibliography}
{\let\endmcitethebibliography\endthebibliography}{}
\begin{mcitethebibliography}{41}
\providecommand*{\natexlab}[1]{#1}
\providecommand*{\mciteSetBstSublistMode}[1]{}
\providecommand*{\mciteSetBstMaxWidthForm}[2]{}
\providecommand*{\mciteBstWouldAddEndPuncttrue}
  {\def\EndOfBibitem{\unskip.}}
\providecommand*{\mciteBstWouldAddEndPunctfalse}
  {\let\EndOfBibitem\relax}
\providecommand*{\mciteSetBstMidEndSepPunct}[3]{}
\providecommand*{\mciteSetBstSublistLabelBeginEnd}[3]{}
\providecommand*{\EndOfBibitem}{}
\mciteSetBstSublistMode{f}
\mciteSetBstMaxWidthForm{subitem}
{(\emph{\alph{mcitesubitemcount}})}
\mciteSetBstSublistLabelBeginEnd{\mcitemaxwidthsubitemform\space}
{\relax}{\relax}

\bibitem[Brown and Maddock(1963)]{Pa}
D.~Brown and A.~G. Maddock, \emph{Q. Rev. Chem. Soc.}, 1963, \textbf{17},
  289--341\relax
\mciteBstWouldAddEndPuncttrue
\mciteSetBstMidEndSepPunct{\mcitedefaultmidpunct}
{\mcitedefaultendpunct}{\mcitedefaultseppunct}\relax
\EndOfBibitem
\bibitem[Wilson(2012)]{WilsonNC}
R.~Wilson, \emph{Nat. Chem.}, 2012, \textbf{4}, 586\relax
\mciteBstWouldAddEndPuncttrue
\mciteSetBstMidEndSepPunct{\mcitedefaultmidpunct}
{\mcitedefaultendpunct}{\mcitedefaultseppunct}\relax
\EndOfBibitem
\bibitem[Wilson \emph{et~al.}(2018)Wilson, De~Sio, and Vallet]{ValletPa}
R.~E. Wilson, S.~De~Sio and V.~Vallet, \emph{Nat. Commun.}, 2018, \textbf{9},
  1--9\relax
\mciteBstWouldAddEndPuncttrue
\mciteSetBstMidEndSepPunct{\mcitedefaultmidpunct}
{\mcitedefaultendpunct}{\mcitedefaultseppunct}\relax
\EndOfBibitem
\bibitem[Le~Naour \emph{et~al.}(2022)Le~Naour, Maloubier, and
  Aupiais]{LeNaour:2022}
C.~Le~Naour, M.~Maloubier and J.~Aupiais, \emph{Radiochim. Acta}, 2022,
  \textbf{110}, 481--493\relax
\mciteBstWouldAddEndPuncttrue
\mciteSetBstMidEndSepPunct{\mcitedefaultmidpunct}
{\mcitedefaultendpunct}{\mcitedefaultseppunct}\relax
\EndOfBibitem
\bibitem[Mendes \emph{et~al.}(2010)Mendes, Hamadi, Le~Naour, Roques, Jeanson,
  Den~Auwer, Moisy, Topin, Aupiais, Hennig, and Di~Giandomenico]{Mendes:2010}
M.~Mendes, S.~Hamadi, C.~Le~Naour, J.~Roques, A.~Jeanson, C.~Den~Auwer,
  P.~Moisy, S.~Topin, J.~Aupiais, C.~Hennig and M.~V. Di~Giandomenico,
  \emph{Inorg. Chem.}, 2010, \textbf{49}, 9962--9971\relax
\mciteBstWouldAddEndPuncttrue
\mciteSetBstMidEndSepPunct{\mcitedefaultmidpunct}
{\mcitedefaultendpunct}{\mcitedefaultseppunct}\relax
\EndOfBibitem
\bibitem[Le~Naour \emph{et~al.}(2005)Le~Naour, Trubert, Di~Giandomenico,
  Fillaux, Den~Auwer, Moisy, and Hennig]{LeNaour:2005}
C.~Le~Naour, D.~Trubert, M.~V. Di~Giandomenico, C.~Fillaux, C.~Den~Auwer,
  P.~Moisy and C.~Hennig, \emph{Inorg. Chem.}, 2005, \textbf{44},
  9542--9546\relax
\mciteBstWouldAddEndPuncttrue
\mciteSetBstMidEndSepPunct{\mcitedefaultmidpunct}
{\mcitedefaultendpunct}{\mcitedefaultseppunct}\relax
\EndOfBibitem
\bibitem[Oher \emph{et~al.}(2023)Oher, Delafoulhouze, Renault, Vallet, and
  Maurice]{PaO-ox3}
H.~Oher, J.~Delafoulhouze, E.~Renault, V.~Vallet and R.~Maurice, \emph{Phys.
  Chem. Chem. Phys.}, 2023, \textbf{25}, 10033--10041\relax
\mciteBstWouldAddEndPuncttrue
\mciteSetBstMidEndSepPunct{\mcitedefaultmidpunct}
{\mcitedefaultendpunct}{\mcitedefaultseppunct}\relax
\EndOfBibitem
\bibitem[Shaaban \emph{et~al.}(2024)Shaaban, Réal, Maurice, and
  Vallet]{Chem-2024}
T.~Shaaban, F.~Réal, R.~Maurice and V.~Vallet, \emph{Chem. Eur. J.}, 2024,
  \textbf{30}, e202304068\relax
\mciteBstWouldAddEndPuncttrue
\mciteSetBstMidEndSepPunct{\mcitedefaultmidpunct}
{\mcitedefaultendpunct}{\mcitedefaultseppunct}\relax
\EndOfBibitem
\bibitem[Brown \emph{et~al.}(1972)Brown, Reynolds, and Moseley]{CrystalS}
D.~Brown, C.~T. Reynolds and P.~T. Moseley, \emph{J. Chem. Soc.{,} Dalton
  Trans.}, 1972,  857--859\relax
\mciteBstWouldAddEndPuncttrue
\mciteSetBstMidEndSepPunct{\mcitedefaultmidpunct}
{\mcitedefaultendpunct}{\mcitedefaultseppunct}\relax
\EndOfBibitem
\bibitem[Kresse and Hafner(1993)]{VASP}
G.~Kresse and J.~Hafner, \emph{Phys. Rev. B}, 1993, \textbf{47}, 558--561\relax
\mciteBstWouldAddEndPuncttrue
\mciteSetBstMidEndSepPunct{\mcitedefaultmidpunct}
{\mcitedefaultendpunct}{\mcitedefaultseppunct}\relax
\EndOfBibitem
\bibitem[Kresse and Joubert(1999)]{PAW}
G.~Kresse and D.~Joubert, \emph{Phys. Rev. B}, 1999, \textbf{59},
  1758--1775\relax
\mciteBstWouldAddEndPuncttrue
\mciteSetBstMidEndSepPunct{\mcitedefaultmidpunct}
{\mcitedefaultendpunct}{\mcitedefaultseppunct}\relax
\EndOfBibitem
\bibitem[Perdew \emph{et~al.}(1997)Perdew, Burke, and Ernzerhof]{PBE}
J.~P. Perdew, K.~Burke and M.~Ernzerhof, \emph{Phys. Rev. Lett.}, 1997,
  \textbf{78}, 1396--1396\relax
\mciteBstWouldAddEndPuncttrue
\mciteSetBstMidEndSepPunct{\mcitedefaultmidpunct}
{\mcitedefaultendpunct}{\mcitedefaultseppunct}\relax
\EndOfBibitem
\bibitem[Dudarev \emph{et~al.}(1998)Dudarev, Botton, Savrasov, Humphreys, and
  Sutton]{U}
S.~L. Dudarev, G.~A. Botton, S.~Y. Savrasov, C.~J. Humphreys and A.~P. Sutton,
  \emph{Phys. Rev. B}, 1998, \textbf{57}, 1505--1509\relax
\mciteBstWouldAddEndPuncttrue
\mciteSetBstMidEndSepPunct{\mcitedefaultmidpunct}
{\mcitedefaultendpunct}{\mcitedefaultseppunct}\relax
\EndOfBibitem
\bibitem[Adamo \emph{et~al.}(1999)Adamo, Scuseria, and Barone]{PBE0}
C.~Adamo, G.~E. Scuseria and V.~Barone, \emph{J. Chem. Phys.}, 1999,
  \textbf{111}, 2889--2899\relax
\mciteBstWouldAddEndPuncttrue
\mciteSetBstMidEndSepPunct{\mcitedefaultmidpunct}
{\mcitedefaultendpunct}{\mcitedefaultseppunct}\relax
\EndOfBibitem
\bibitem[Grimme \emph{et~al.}(2011)Grimme, Ehrlich, and Goerigk]{D3BJ}
S.~Grimme, S.~Ehrlich and L.~Goerigk, \emph{J. Comput. Chem.}, 2011,
  \textbf{32}, 1456--1465\relax
\mciteBstWouldAddEndPuncttrue
\mciteSetBstMidEndSepPunct{\mcitedefaultmidpunct}
{\mcitedefaultendpunct}{\mcitedefaultseppunct}\relax
\EndOfBibitem
\bibitem[Rajapaksha \emph{et~al.}(2023)Rajapaksha, Augustine, Mason, and
  Forbes]{AngewS}
H.~Rajapaksha, L.~J. Augustine, S.~E. Mason and T.~Z. Forbes, \emph{Angew.
  Chem. Int. Ed.}, 2023, \textbf{62}, e202305073\relax
\mciteBstWouldAddEndPuncttrue
\mciteSetBstMidEndSepPunct{\mcitedefaultmidpunct}
{\mcitedefaultendpunct}{\mcitedefaultseppunct}\relax
\EndOfBibitem
\bibitem[Rajapaksha \emph{et~al.}(2023)Rajapaksha, Mason, and Forbes]{ICS}
H.~Rajapaksha, S.~E. Mason and T.~Z. Forbes, \emph{Inorg. Chem.}, 2023,
  \textbf{62}, 14318--14325\relax
\mciteBstWouldAddEndPuncttrue
\mciteSetBstMidEndSepPunct{\mcitedefaultmidpunct}
{\mcitedefaultendpunct}{\mcitedefaultseppunct}\relax
\EndOfBibitem
\bibitem[Müller \emph{et~al.}(2021)Müller, Ertural, Hempelmann, and
  Dronskowski]{ICOBIs}
P.~C. Müller, C.~Ertural, J.~Hempelmann and R.~Dronskowski, \emph{J. Phys.
  Chem. C}, 2021, \textbf{125}, 7959--7970\relax
\mciteBstWouldAddEndPuncttrue
\mciteSetBstMidEndSepPunct{\mcitedefaultmidpunct}
{\mcitedefaultendpunct}{\mcitedefaultseppunct}\relax
\EndOfBibitem
\bibitem[Momma and Izumi(2008)]{Vesta}
K.~Momma and F.~Izumi, \emph{J. Appl. Crystallogr.}, 2008, \textbf{41},
  653--658\relax
\mciteBstWouldAddEndPuncttrue
\mciteSetBstMidEndSepPunct{\mcitedefaultmidpunct}
{\mcitedefaultendpunct}{\mcitedefaultseppunct}\relax
\EndOfBibitem
\bibitem[Gonze \emph{et~al.}(2020)Gonze, Amadon, Antonius, Arnardi, Baguet,
  Beuken, Bieder, Bottin, Bouchet, Bousquet, Brouwer, Bruneval, Brunin,
  Cavignac, Charraud, Chen, Côté, Cottenier, Denier, Geneste, Ghosez,
  Giantomassi, Gillet, Gingras, Hamann, Hautier, He, Helbig, Holzwarth, Jia,
  Jollet, Lafargue-Dit-Hauret, Lejaeghere, Marques, Martin, Martins, Miranda,
  Naccarato, Persson, Petretto, Planes, Pouillon, Prokhorenko, Ricci,
  Rignanese, Romero, Schmitt, Torrent, {van Setten}, {Van Troeye}, Verstraete,
  Zérah, and Zwanziger]{Abinit}
X.~Gonze, B.~Amadon, G.~Antonius, F.~Arnardi, L.~Baguet, J.-M. Beuken,
  J.~Bieder, F.~Bottin, J.~Bouchet, E.~Bousquet, N.~Brouwer, F.~Bruneval,
  G.~Brunin, T.~Cavignac, J.-B. Charraud, W.~Chen, M.~Côté, S.~Cottenier,
  J.~Denier, G.~Geneste, P.~Ghosez, M.~Giantomassi, Y.~Gillet, O.~Gingras,
  D.~R. Hamann, G.~Hautier, X.~He, N.~Helbig, N.~Holzwarth, Y.~Jia, F.~Jollet,
  W.~Lafargue-Dit-Hauret, K.~Lejaeghere, M.~A. Marques, A.~Martin, C.~Martins,
  H.~P. Miranda, F.~Naccarato, K.~Persson, G.~Petretto, V.~Planes, Y.~Pouillon,
  S.~Prokhorenko, F.~Ricci, G.-M. Rignanese, A.~H. Romero, M.~M. Schmitt,
  M.~Torrent, M.~J. {van Setten}, B.~{Van Troeye}, M.~J. Verstraete, G.~Zérah
  and J.~W. Zwanziger, \emph{Comput. Phys. Commun.}, 2020, \textbf{248},
  107042\relax
\mciteBstWouldAddEndPuncttrue
\mciteSetBstMidEndSepPunct{\mcitedefaultmidpunct}
{\mcitedefaultendpunct}{\mcitedefaultseppunct}\relax
\EndOfBibitem
\bibitem[Siberchicot \emph{et~al.}(2021)Siberchicot, Aupiais, and
  Naour]{Fluoro}
B.~Siberchicot, J.~Aupiais and C.~L. Naour, \emph{Radiochim. Acta}, 2021,
  \textbf{109}, 673--680\relax
\mciteBstWouldAddEndPuncttrue
\mciteSetBstMidEndSepPunct{\mcitedefaultmidpunct}
{\mcitedefaultendpunct}{\mcitedefaultseppunct}\relax
\EndOfBibitem
\bibitem[Bukhsh \emph{et~al.}(1966)Bukhsh, Flegenheimer, Hall, Maddock, and {de
  Miranda}]{PaF8}
M.~Bukhsh, J.~Flegenheimer, F.~Hall, A.~Maddock and C.~{de Miranda}, \emph{J.
  Inorg. Nucl. Chem.}, 1966, \textbf{28}, 421--431\relax
\mciteBstWouldAddEndPuncttrue
\mciteSetBstMidEndSepPunct{\mcitedefaultmidpunct}
{\mcitedefaultendpunct}{\mcitedefaultseppunct}\relax
\EndOfBibitem
\bibitem[Frisch \emph{et~al.}(2016)Frisch, Trucks, Schlegel, Scuseria, Robb,
  Cheeseman, Scalmani, Barone, Petersson, Nakatsuji, Li, Caricato, Marenich,
  Bloino, Janesko, Gomperts, Mennucci, Hratchian, Ortiz, Izmaylov, Sonnenberg,
  Williams-Young, Ding, Lipparini, Egidi, Goings, Peng, Petrone, Henderson,
  Ranasinghe, Zakrzewski, Gao, Rega, Zheng, Liang, Hada, Ehara, Toyota, Fukuda,
  Hasegawa, Ishida, Nakajima, Honda, Kitao, Nakai, Vreven, Throssell,
  Montgomery, Peralta, Ogliaro, Bearpark, Heyd, Brothers, Kudin, Staroverov,
  Keith, Kobayashi, Normand, Raghavachari, Rendell, Burant, Iyengar, Tomasi,
  Cossi, Millam, Klene, Adamo, Cammi, Ochterski, Martin, Morokuma, Farkas,
  Foresman, and Fox]{Gaussian}
M.~J. Frisch, G.~W. Trucks, H.~B. Schlegel, G.~E. Scuseria, M.~A. Robb, J.~R.
  Cheeseman, G.~Scalmani, V.~Barone, G.~A. Petersson, H.~Nakatsuji, X.~Li,
  M.~Caricato, A.~V. Marenich, J.~Bloino, B.~G. Janesko, R.~Gomperts,
  B.~Mennucci, H.~P. Hratchian, J.~V. Ortiz, A.~F. Izmaylov, J.~L. Sonnenberg,
  D.~Williams-Young, F.~Ding, F.~Lipparini, F.~Egidi, J.~Goings, B.~Peng,
  A.~Petrone, T.~Henderson, D.~Ranasinghe, V.~G. Zakrzewski, J.~Gao, N.~Rega,
  G.~Zheng, W.~Liang, M.~Hada, M.~Ehara, K.~Toyota, R.~Fukuda, J.~Hasegawa,
  M.~Ishida, T.~Nakajima, Y.~Honda, O.~Kitao, H.~Nakai, T.~Vreven,
  K.~Throssell, J.~A. Montgomery, {Jr.}, J.~E. Peralta, F.~Ogliaro, M.~J.
  Bearpark, J.~J. Heyd, E.~N. Brothers, K.~N. Kudin, V.~N. Staroverov, T.~A.
  Keith, R.~Kobayashi, J.~Normand, K.~Raghavachari, A.~P. Rendell, J.~C.
  Burant, S.~S. Iyengar, J.~Tomasi, M.~Cossi, J.~M. Millam, M.~Klene, C.~Adamo,
  R.~Cammi, J.~W. Ochterski, R.~L. Martin, K.~Morokuma, O.~Farkas, J.~B.
  Foresman and D.~J. Fox, \emph{Gaussian˜16 {R}evision {C}.01}, 2016, Gaussian
  Inc. Wallingford CT\relax
\mciteBstWouldAddEndPuncttrue
\mciteSetBstMidEndSepPunct{\mcitedefaultmidpunct}
{\mcitedefaultendpunct}{\mcitedefaultseppunct}\relax
\EndOfBibitem
\bibitem[Werner \emph{et~al.}()Werner, Knowles, Knizia, Manby, {Sch\"{u}tz},
  Celani, Gy\"orffy, Kats, Korona, Lindh, Mitrushenkov, Rauhut, Shamasundar,
  Adler, Amos, Bennie, Bernhardsson, Berning, Cooper, Deegan, Dobbyn, Eckert,
  Goll, Hampel, Hesselmann, Hetzer, Hrenar, Jansen, K\"oppl, Lee, Liu, Lloyd,
  Ma, Mata, May, McNicholas, Meyer, {Miller III}, Mura, Nicklass, O'Neill,
  Palmieri, Peng, Pfl\"uger, Pitzer, Reiher, Shiozaki, Stoll, Stone, Tarroni,
  Thorsteinsson, Wang, and Welborn]{Molpro}
H.-J. Werner, P.~J. Knowles, G.~Knizia, F.~R. Manby, M.~{Sch\"{u}tz},
  P.~Celani, W.~Gy\"orffy, D.~Kats, T.~Korona, R.~Lindh, A.~Mitrushenkov,
  G.~Rauhut, K.~R. Shamasundar, T.~B. Adler, R.~D. Amos, S.~J. Bennie,
  A.~Bernhardsson, A.~Berning, D.~L. Cooper, M.~J.~O. Deegan, A.~J. Dobbyn,
  F.~Eckert, E.~Goll, C.~Hampel, A.~Hesselmann, G.~Hetzer, T.~Hrenar,
  G.~Jansen, C.~K\"oppl, S.~J.~R. Lee, Y.~Liu, A.~W. Lloyd, Q.~Ma, R.~A. Mata,
  A.~J. May, S.~J. McNicholas, W.~Meyer, T.~F. {Miller III}, M.~E. Mura,
  A.~Nicklass, D.~P. O'Neill, P.~Palmieri, D.~Peng, K.~Pfl\"uger, R.~Pitzer,
  M.~Reiher, T.~Shiozaki, H.~Stoll, A.~J. Stone, R.~Tarroni, T.~Thorsteinsson,
  M.~Wang and M.~Welborn, \emph{MOLPRO, a package of ab initio programs,
  version 2023.2}\relax
\mciteBstWouldAddEndPuncttrue
\mciteSetBstMidEndSepPunct{\mcitedefaultmidpunct}
{\mcitedefaultendpunct}{\mcitedefaultseppunct}\relax
\EndOfBibitem
\bibitem[te~Velde \emph{et~al.}(2001)te~Velde, Bickelhaupt, Baerends,
  Fonseca~Guerra, van Gisbergen, Snijders, and Ziegler]{ADF}
G.~te~Velde, F.~M. Bickelhaupt, E.~J. Baerends, C.~Fonseca~Guerra, S.~J.~A. van
  Gisbergen, J.~G. Snijders and T.~Ziegler, \emph{J. Comput. Chem.}, 2001,
  \textbf{22}, 931--967\relax
\mciteBstWouldAddEndPuncttrue
\mciteSetBstMidEndSepPunct{\mcitedefaultmidpunct}
{\mcitedefaultendpunct}{\mcitedefaultseppunct}\relax
\EndOfBibitem
\bibitem[Stephens \emph{et~al.}(1994)Stephens, Devlin, Chabalowski, and
  Frisch]{B3LYP}
P.~J. Stephens, F.~J. Devlin, C.~F. Chabalowski and M.~J. Frisch, \emph{J.
  Chem. Phys.}, 1994, \textbf{98}, 11623--11627\relax
\mciteBstWouldAddEndPuncttrue
\mciteSetBstMidEndSepPunct{\mcitedefaultmidpunct}
{\mcitedefaultendpunct}{\mcitedefaultseppunct}\relax
\EndOfBibitem
\bibitem[Purvis and Bartlett(1982)]{CCSD(T)}
I.~Purvis, George~D. and R.~J. Bartlett, \emph{J. Chem. Phys.}, 1982,
  \textbf{76}, 1910--1918\relax
\mciteBstWouldAddEndPuncttrue
\mciteSetBstMidEndSepPunct{\mcitedefaultmidpunct}
{\mcitedefaultendpunct}{\mcitedefaultseppunct}\relax
\EndOfBibitem
\bibitem[M\o{}ller and Plesset(1934)]{MP2}
C.~M\o{}ller and M.~S. Plesset, \emph{Phys. Rev.}, 1934, \textbf{46},
  618--622\relax
\mciteBstWouldAddEndPuncttrue
\mciteSetBstMidEndSepPunct{\mcitedefaultmidpunct}
{\mcitedefaultendpunct}{\mcitedefaultseppunct}\relax
\EndOfBibitem
\bibitem[Pye and Ziegler(1999)]{COSMO-ADF}
C.~C. Pye and T.~Ziegler, \emph{Theor. Chem. Acc.}, 1999, \textbf{101},
  396--408\relax
\mciteBstWouldAddEndPuncttrue
\mciteSetBstMidEndSepPunct{\mcitedefaultmidpunct}
{\mcitedefaultendpunct}{\mcitedefaultseppunct}\relax
\EndOfBibitem
\bibitem[Outeiral \emph{et~al.}(2018)Outeiral, Vincent, Mart\'in~Pend\'as, and
  Popelier]{DI}
C.~Outeiral, M.~A. Vincent, A.~Mart\'in~Pend\'as and P.~L.~A. Popelier,
  \emph{Chem. Sci.}, 2018, \textbf{9}, 5517--5529\relax
\mciteBstWouldAddEndPuncttrue
\mciteSetBstMidEndSepPunct{\mcitedefaultmidpunct}
{\mcitedefaultendpunct}{\mcitedefaultseppunct}\relax
\EndOfBibitem
\bibitem[Bader(1991)]{QTAIM}
R.~F.~W. Bader, \emph{Chem. Rev.}, 1991, \textbf{91}, 893--928\relax
\mciteBstWouldAddEndPuncttrue
\mciteSetBstMidEndSepPunct{\mcitedefaultmidpunct}
{\mcitedefaultendpunct}{\mcitedefaultseppunct}\relax
\EndOfBibitem
\bibitem[Vallet \emph{et~al.}(2003)Vallet, Moll, Wahlgren, Szabó, and
  Grenthe]{UO2-ox3}
V.~Vallet, H.~Moll, U.~Wahlgren, Z.~Szabó and I.~Grenthe, \emph{Inorg. Chem.},
  2003, \textbf{42}, 8598--8598\relax
\mciteBstWouldAddEndPuncttrue
\mciteSetBstMidEndSepPunct{\mcitedefaultmidpunct}
{\mcitedefaultendpunct}{\mcitedefaultseppunct}\relax
\EndOfBibitem
\bibitem[{de Jong} \emph{et~al.}(1999){de Jong}, Visscher, and
  Nieuwpoort]{deJong:1999}
W.~A. {de Jong}, L.~Visscher and W.~C. Nieuwpoort, \emph{J. Mol. Struct.
  THEOCHEM}, 1999, \textbf{458}, 41--52\relax
\mciteBstWouldAddEndPuncttrue
\mciteSetBstMidEndSepPunct{\mcitedefaultmidpunct}
{\mcitedefaultendpunct}{\mcitedefaultseppunct}\relax
\EndOfBibitem
\bibitem[Servaes \emph{et~al.}(2005)Servaes, Hennig, Van~Deun, and
  Görller-Walrand]{UO2Cl4}
K.~Servaes, C.~Hennig, R.~Van~Deun and C.~Görller-Walrand, \emph{Inorg.
  Chem.}, 2005, \textbf{44}, 7705--7707\relax
\mciteBstWouldAddEndPuncttrue
\mciteSetBstMidEndSepPunct{\mcitedefaultmidpunct}
{\mcitedefaultendpunct}{\mcitedefaultseppunct}\relax
\EndOfBibitem
\bibitem[Denning(2007)]{Denning}
R.~G. Denning, \emph{J. Phys. Chem. A}, 2007, \textbf{111}, 4125--4143\relax
\mciteBstWouldAddEndPuncttrue
\mciteSetBstMidEndSepPunct{\mcitedefaultmidpunct}
{\mcitedefaultendpunct}{\mcitedefaultseppunct}\relax
\EndOfBibitem
\bibitem[Wellington \emph{et~al.}(2016)Wellington, Kerridge, and
  Kaltsoyannis]{PolarBond}
J.~P. Wellington, A.~Kerridge and N.~Kaltsoyannis, \emph{Polyhedron}, 2016,
  \textbf{116}, 57--63\relax
\mciteBstWouldAddEndPuncttrue
\mciteSetBstMidEndSepPunct{\mcitedefaultmidpunct}
{\mcitedefaultendpunct}{\mcitedefaultseppunct}\relax
\EndOfBibitem
\bibitem[Pershina(1999)]{actinide-Pershina-RA1999-84-79--84}
V.~Pershina, \emph{Radiochim. Acta}, 1999, \textbf{84}, 79--84\relax
\mciteBstWouldAddEndPuncttrue
\mciteSetBstMidEndSepPunct{\mcitedefaultmidpunct}
{\mcitedefaultendpunct}{\mcitedefaultseppunct}\relax
\EndOfBibitem
\bibitem[Pershina(1998)]{actinide-Pershina-RA1998-80-65--73}
V.~Pershina, \emph{Radiochim. Acta}, 1998, \textbf{80}, 65--73\relax
\mciteBstWouldAddEndPuncttrue
\mciteSetBstMidEndSepPunct{\mcitedefaultmidpunct}
{\mcitedefaultendpunct}{\mcitedefaultseppunct}\relax
\EndOfBibitem
\bibitem[O'Grady and Kaltsoyannis(2002)]{actinide-OGrady-JCDT2002--1233--1239}
E.~O'Grady and N.~Kaltsoyannis, \emph{J. Chem. Soc., Dalton Trans.}, 2002,
  1233--1239\relax
\mciteBstWouldAddEndPuncttrue
\mciteSetBstMidEndSepPunct{\mcitedefaultmidpunct}
{\mcitedefaultendpunct}{\mcitedefaultseppunct}\relax
\EndOfBibitem
\bibitem[{Fryer-Kanssen} and
  Kerridge(2018)]{Fryer-Kanssen-Chem.Commun.-2018-54-9761-9764}
I.~{Fryer-Kanssen} and A.~Kerridge, \emph{Chem. Commun.}, 2018, \textbf{54},
  9761--9764\relax
\mciteBstWouldAddEndPuncttrue
\mciteSetBstMidEndSepPunct{\mcitedefaultmidpunct}
{\mcitedefaultendpunct}{\mcitedefaultseppunct}\relax
\EndOfBibitem
\bibitem[Motta and Autschbach(2023)]{actinide-Motta-NC2023-14-4307}
L.~C. Motta and J.~Autschbach, \emph{Nat. Chem.}, 2023, \textbf{14}, 4307\relax
\mciteBstWouldAddEndPuncttrue
\mciteSetBstMidEndSepPunct{\mcitedefaultmidpunct}
{\mcitedefaultendpunct}{\mcitedefaultseppunct}\relax
\EndOfBibitem
\end{mcitethebibliography}
\clearpage

\includepdf[pages=1-15]{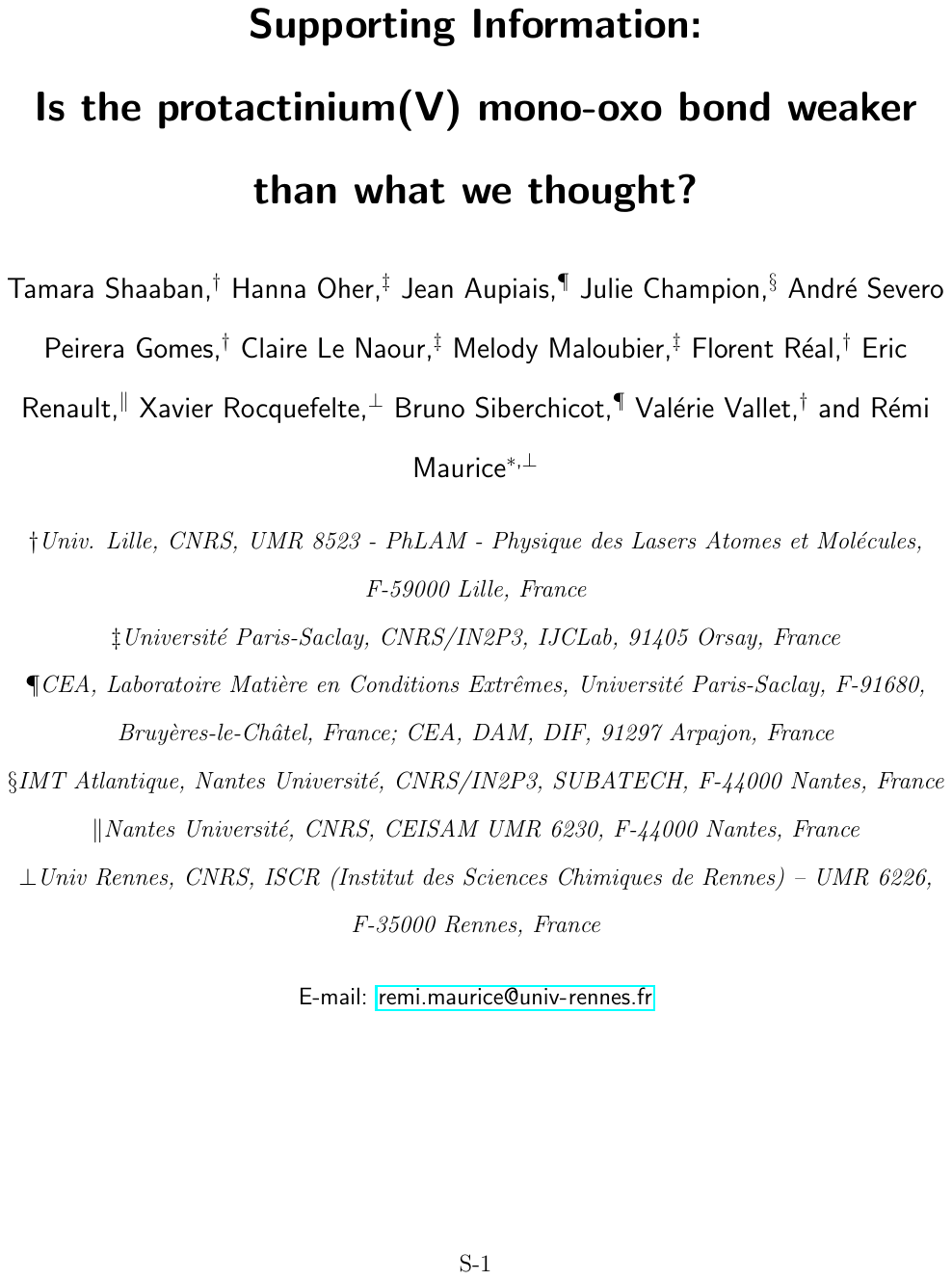}
\end{document}